\documentclass[12pt,a4paper]{article}
\usepackage{times}
\usepackage{a4wide}
\usepackage{amsfonts}
\usepackage{amssymb}
\usepackage{amsmath}
\usepackage{ifpdf}
\ifpdf
\usepackage[pdftex,unicode,implicit]{hyperref}
\hypersetup{%
  pdftitle    = {Quantum black holes in Type-IIA String Theory},
  pdfkeywords = {string theory, black hole, Calabi Yau, compactification, quantum corrections, supergravity model},
  pdfauthor   = {Pablo Bueno, Rhys Davies, and C. S. Shahbazi},
  plainpages  = true,
  colorlinks  = true,
  citecolor   = blue,
  urlcolor    = red,
  linkcolor   = black
}

\newcommand{\arxiv}[1]{{\tt
\href{http://www.arXiv.org/abs/#1}{arXiv:#1}}}
\else
  \usepackage[dvips]{graphicx}
  \usepackage[unicode,implicit]{hyperref}

  \newcommand{\arxiv}[1]{{\tt arXiv:#1}}
\fi
\makeatletter
\@addtoreset{equation}{section}
\makeatother

\usepackage{amssymb,amsmath}

\renewcommand{\a}{\alpha}

\renewcommand{\d}{\delta}

\renewcommand{\k}{\kappa}

\newcommand{\s}{\sigma}

\newcommand{\IP}{\mathbb{P}}

\newcommand{\IZ}{\mathbb{Z}}

\newcommand{\cicy}[2]{\begin{matrix} #1\end{matrix}\!\left[\begin{matrix}#2 \end{matrix}\right]}
\newcommand{\quotient}[1]{_{\hskip-2pt\lower1pt\hbox{$/$}\lower2pt\hbox{\hskip-1pt$#1$}}}

\def\place#1#2#3{\vbox to0pt{\kern-\parskip\kern-7pt
                             \kern-#2truein\hbox{\kern#1truein #3}
                             \vss}\nointerlineskip}

\newcommand{\rest}[1]{\big\vert_{#1}}

\newcommand{\Dic}{\mathrm{Dic}}

\newcommand{\hodgenos}{(h^{1,1},\,h^{2,1})}
\newcommand{\symm}[1]{_{\hskip-3pt\lower3pt\hbox{$\left\{#1\right\}$}}}

\newcommand{\cicystop}{~\lower8pt\hbox{.}}

\def\place#1#2#3{\vbox to0pt{\kern-\parskip\kern-7pt
                             \kern-#2truein\hbox{\kern#1truein #3}
                             \vss}\nointerlineskip}

\newcommand{\beq}{\begin{equation}}
\newcommand{\eeq}{\end{equation}}
\newcommand{\bea}{\begin{eqnarray}}
\newcommand{\eea}{\end{eqnarray}}
\newcommand{\bean}{\begin{eqnarray*}}
\newcommand{\eean}{\end{eqnarray*}}

\newcommand{\comment}[1]{}

\begin{document}

\begin{flushright}
\small
IFT-UAM/CSIC-12-91\\
%\texttt{arXiv:yymm.nnnn [hep-th]}\\
October 9\textsuperscript{th}, 2012\\
\normalsize
\end{flushright}

\begin{center}

\vspace{1cm}

{\LARGE {\bf Quantum black holes in Type-IIA\\[8mm]  String Theory}}

\vspace{2.5cm}

\begin{center}

\renewcommand{\thefootnote}{\alph{footnote}}
{\sl\large Pablo~Bueno$^{\heartsuit}$}
\footnote{E-mail: {\tt pab.bueno [at] estudiante.uam.es}},
{\sl\large Rhys~Davies$^{\sharp}$}
\footnote{E-mail: {\tt daviesr [at] maths.ox.ac.uk}}
{\sl\large and C.~S.~Shahbazi$^{\diamondsuit\heartsuit}$}
\footnote{E-mail: {\tt Carlos.Shabazi [at] uam.es}}
\renewcommand{\thefootnote}{\arabic{footnote}}

\vspace{.4cm}

${}^{\heartsuit}${\it Instituto de F\'{\i}sica Te\'orica UAM/CSIC\\
C/ Nicol\'as Cabrera, 13--15,  C.U.~Cantoblanco, 28049 Madrid, Spain}\\

\vspace{.2cm}

${}^\sharp${\it
Mathematical Institute, University of Oxford, \\
24-29 St Giles, Oxford, OX1 3LB, UK}

\vspace{.2cm}

${}^{\diamondsuit}${\it Stanford Institute for Theoretical Physics
and Department of Physics, Stanford University,\\
  Stanford, CA 94305-4060, USA}\\

\vspace{1.5cm}

\end{center}

{\bf Abstract}

\begin{quotation}

  {\small
We study black hole solutions of Type-IIA Calabi--Yau compactifications in the presence of quantum perturbative corrections. We define a class of black holes that only exist in the presence of quantum corrections and that, consequently, can be considered as purely \emph{quantum} black holes. The regularity conditions of the solutions impose the topological constraint $h^{1,1}>h^{2,1}$ on the Calabi--Yau manifold, defining a class of admissible compactifications, which we prove to be non-empty for $h^{1,1}=3$ by explicitly constructing the corresponding Calabi--Yau manifolds, new in the literature.

}

\end{quotation}

\end{center}

\setcounter{footnote}{0}

\newpage
%%%%%%%%%%%%%%%%%%%%%%%%%%%%%%%%%%%%%%%%%%%%%%%%%%%%%%%%%%%%%%%%%%%%%%
%%%%%%%%%%%%%%%%%%%%%%%%%%%%%%%%%%%%%%%%%%%%%%%%%%%%%%%%%%%%%%%%%%%%%%
%%%%%%%%%%%%%%%%%%%%%%%%%%%%%%%%%%%%%%%%%%%%%%%%%%%%%%%%%%%%%%%%%%%%%%
%%%%%%%%%%%%%%%%%%%%%%%%%%%%%%%%%%%%%%%%%%%%%%%%%%%%%%%%%%%%%%%%%%%%%%
\pagestyle{plain}
%%%%%%%%%%%%%%%%%%%%%%%%%%%%%%%%%%%%%%%%%%%%%%%%%%%%%%%%%%%%%%%%%%%%%%
%%%%%%%%%%%%%%%%%%%%%%%%%%%%%%%%%%%%%%%%%%%%%%%%%%%%%%%%%%%%%%%%%%%%%%
%%%%%%%%%%%%%%%%%%%%%%%%%%%%%%%%%%%%%%%%%%%%%%%%%%%%%%%%%%%%%%%%%%%%%%
%%%%%%%%%%%%%%%%%%%%%%%%%%%%%%%%%%%%%%%%%%%%%%%%%%%%%%%%%%%%%%%%%%%%%%
%%%%%%%%%%%%%%%%%%%%%%%%%%%%%%%%%%%%%%%%%%%%%%%%%%%%%%%%%%%%%%%%%%%%%%

\tableofcontents

%%%%%%%%%%%%%%%%%%%%%%%%%%%%%%%%%%%%%%%%%%%%%%%%%%%%%%%%%%%%%%%%%%%%%%
%%%%%%%%%%%%%%%%%%%%%%%%%%%%%%%%%%%%%%%%%%%%%%%%%%%%%%%%%%%%%%%%%%%%%%
%%%%%%%%%%%%%%%%%%%%%%%%%%%%%%%%%%%%%%%%%%%%%%%%%%%%%%%%%%%%%%%%%%%%%%
%%%%%%%%%%%%%%%%%%%%%%%%%%%%%%%%%%%%%%%%%%%%%%%%%%%%%%%%%%%%%%%%%%%%%%

\section*{General remarks}

Supergravity solutions have played, and continue to play, a prominent r\^ole in the new developments of String Theory. The body of literature about black hole solutions (and $p$-branes) that has been accumulated during the past thirty years is enormous, but only recently the issue of non-extremality was systematically investigated, and by now, it could be said that we have at our disposal well-established methods to deal with non-extremal solutions \cite{Chemissany:2010zp,Chemissany:2010ay,Mohaupt:2010fk,Mohaupt:2011aa,Klemm:2012yg,deAntonioMartin:2012bi,Galli:2011fq,Meessen:2011aa,Meessen:2012su} in  Supergravity.  However, explicit non-extremal solutions to Supergravity models with perturbative quantum corrections are yet to be constructed. These  kind of solutions may be relevant in order to understand how the deformation of the scalar geometry modifies the solutions of the theory, and also in order to relate the macroscopic computation of the entropy with the microscopic calculation in a String Theory 
set-up, once sub-leading corrections to the prepotential are taken into account \cite{Behrndt:1997gs,Behrndt:1997ei}. These kinds of corrections differ from the higher order corrections, which, together with the corresponding microscopic String Theory computation, have been extensively studied in the literature \cite{Lopes Cardoso:1999ur} (for a very nice review about this and related topics, as well as for further references, see \cite{Mohaupt:2000mj}).

In this note we are going to use the so-called \emph{H-formalism} \cite{Galli:2011fq,Meessen:2011aa,Meessen:2012su} in order to take a small step in the study of non-extremal black holes in Supergravity in the presence of quantum corrections. The \emph{H-formalism}, as it has been used so far to produce new solutions, is based on a change of variables in the $\mathcal{N}=2,~d=4$ ungauged Supergravity action (to new ones $H^M$ that transform linearly under duality and become harmonic functions on $\mathbb{R}^{3}$ in the extremal case) plus a hyperbolic Ansatz for them, that allows to transform the system of differential equations of motion into a system of algebraic equations, easier to handle. It is, of course, also possible to make the change of variables and try to solve the resulting system of differential equations by other means, not involving any particular Ansatz for the $H^{M}$.

Through a consistent truncation, we are going to define a particular class of black holes, which is characterized by existing only when the quantum perturbative corrections are included in the action. These kinds of solutions, which we have chosen to call \emph{quantum black holes}, display a remarkable behavior: the so called~ \emph{large-volume limit} $\Im{\rm m}z^{i}\rightarrow \infty$ is in fact not a large volume limit of the Calabi-Yau (C.Y.) manifold, whose volume remains constant and fixed by topological data. In addition, the regularity conditions of the black hole solutions impose the topological restriction $h^{1,1}>h^{2,1}$ in the compactification C.Y. For small $h^{1,1}$ the condition is particularly restrictive, and since this case is the most manageable one from the point of view of black hole solutions, we prove the existence of C.Y. manifolds obeying $h^{1,1} >h^{2,1}$ by explicit construction for the $h^{1,1}=3$ case. These C.Y. manifolds are new in the literature.

The perturbative corrections, encoded in a single term $~i\frac{c}{2}$ in the prepotential, introduce a highly non-trivial difficulty in the model, which makes almost hopeless the resolution of the equations of the theory. Surprisingly enough, we are able to find a black hole solution with non-constant scalars, similar to the $D0-D4-D4-D4$ black hole solution of the $STU$ model, and which can be used as a toy model to study the microscopic description of black holes in the presence of quantum perturbative corrections and away from extremality.

%%%%%%%%%%%%%%%%%%%%%%%%%%%%%%%%%%%%%%%%%%%%%%%%%%%%%%%%%%%%%%%%%%%%%%
%%%%%%%%%%%%%%%%%%%%%%%%%%%%%%%%%%%%%%%%%%%%%%%%%%%%%%%%%%%%%%%%%%%%%%
%%%%%%%%%%%%%%%%%%%%%%%%%%%%%%%%%%%%%%%%%%%%%%%%%%%%%%%%%%%%%%%%%%%%%%
%%%%%%%%%%%%%%%%%%%%%%%%%%%%%%%%%%%%%%%%%%%%%%%%%%%%%%%%%%%%%%%%%%%%%%

%%%%%%%%%%%%%%%%%%%%%%%%%%%%%%%%%%%%%%%%%%%%%%%%%%%%%%%%%%%%%%%%%%%%%%
%%%%%%%%%%%%%%%%%%%%%%%%%%%%%%%%%%%%%%%%%%%%%%%%%%%%%%%%%%%%%%%%%%%%%%
%%%%%%%%%%%%%%%%%%%%%%%%%%%%%%%%%%%%%%%%%%%%%%%%%%%%%%%%%%%%%%%%%%%%%%
%%%%%%%%%%%%%%%%%%%%%%%%%%%%%%%%%%%%%%%%%%%%%%%%%%%%%%%%%%%%%%%%%%%%%%

\section{Type-IIA String Theory on a Calabi--Yau manifold}
\label{sec:generaltheory}

%%%%%%%%%%%%%%%%%%%%%%%%%%%%%%%%%%%%%%%%%%%%%%%%%%%%%%%%%%%%%%%%%%%%%%
%%%%%%%%%%%%%%%%%%%%%%%%%%%%%%%%%%%%%%%%%%%%%%%%%%%%%%%%%%%%%%%%%%%%%%
%%%%%%%%%%%%%%%%%%%%%%%%%%%%%%%%%%%%%%%%%%%%%%%%%%%%%%%%%%%%%%%%%%%%%%
%%%%%%%%%%%%%%%%%%%%%%%%%%%%%%%%%%%%%%%%%%%%%%%%%%%%%%%%%%%%%%%%%%%%%%

Type-IIA String Theory compactified to four dimensions on a C.Y. manifold, with Hodge numbers $(h^{1,1},h^{2,1})$, is described by a $\mathcal{N}=2, d=4$ Supergravity whose prepotential is given in terms of an infinite series around $\Im{\rm m}z^i\rightarrow \infty$\footnote{Actually, the prepotential obtained in a Type-IIA C.Y. compactification is \emph{symplectically equivalent} to the prepotential (\ref{eq:IIaprepotential}).} \cite{Candelas:1990pi,Candelas:1990rm,Candelas:1990qd}

\begin{equation}
\label{eq:IIaprepotential}
\mathcal{F} =  -\frac{1}{3!}\kappa^{0}_{ijk} z^i z^j z^k +\frac{ic}{2}+\frac{i}{(2\pi)^3}\sum_{d_{i}} n_{d_{i}} Li_{3}\left(e^{2\pi i d_{i} z^{i}}\right) \ \, ,
\end{equation}

\noindent
where $z^i,~~ i =1,...,n_v=h^{1,1}$, are the scalars in the vector multiplets,\footnote{There are also $h^{2,1}+1$ hypermultiplets in the theory. However, they can be consistently set to a constant value.} $c=\frac{\chi\zeta(3)}{ (2\pi)^3}$ is a model dependent number\footnote{$\chi$ is the Euler characteristic, which for C.Y. three-folds is given by $\chi=2(h^{1,1}-h^{2,1})$.}, $\kappa^{0}_{ijk}$ are the classical intersection numbers, $d_{i}\in\mathbb{Z}^{+}$ is $n_{v}$-dimensional summation index and

\begin{equation}
\label{eq:Li}
Li_{3}(x)=\sum^{\infty}_{j=1}\frac{x^j}{j^3} \ \, .
\end{equation}

\noindent
The non-perturbative part of the prepotential (\ref{eq:IIaprepotential}), according to his stringy origin, is given by

\begin{equation}
\label{eq:IIanonpert}
\mathcal{F}_{Non-Pert} = \frac{i}{(2\pi)^3}\sum_{d_{i}} n_{d_{i}} Li_{3}\left(e^{2\pi i d_{i} z^{i}}\right) \ \, ,
\end{equation}

\noindent
whereas the rest of the prepotential includes the tree level contribution and the quantum perturbative corrections

\begin{equation}
\label{eq:IIapert}
\mathcal{F}_{Pert} =  -\frac{1}{3!}\kappa^{0}_{ijk} z^i z^j z^k +\frac{ic}{2}\ \, .
\end{equation}

\noindent
The non-perturbative corrections (\ref{eq:IIanonpert}) are exponentially suppressed and therefore can be safely ignored going to the large volume limit. Therefore our starting point is going to be Eq. (\ref{eq:IIapert}), which in homogeneous coordinates $\mathcal{X}^{\Lambda}, ~~\Lambda=(0,i)$, can be written as

\begin{equation}
\label{eq:prepotential}
F(\mathcal{X})=  - \frac{1}{3!}\kappa^{0}_{ijk}  \frac{\mathcal{X}^{i}\mathcal{X}^{j}\mathcal{X}^{k}}{\mathcal{X}^0} +\frac{ic}{2} \left(\mathcal{X}^0\right)^2 \ \, .
\end{equation}

\noindent
The scalars $z^i$ are given by\footnote{This coordinate system is therefore only valid away from the locus $\mathcal{X}^{0}=0$. }

\begin{equation}
\label{eq:scalarsgeneral}
z^i=\frac{\mathcal{X}^{i}}{\mathcal{X}^0} \ \, .
\end{equation}

\noindent
The scalar geometry defined by (\ref{eq:prepotential}) is the so called \emph{quantum corrected} $d$-SK geometry\footnote{The attractor points of this model have been extensively studied in \cite{Bellucci:2009pg}. Related works can be found in \cite{Bellucci:2007eh} \cite{Bellucci:2008tx}.}  \cite{deWit:1991nm}, \cite{deWit:1992wf}. In this scenario, the classical case is modified and the scalar manifold, due to the correction encoded in $c$,  is no longer homogeneous, and therefore, the geometry has been \emph{corrected} by quantum effects.

We are interested in studying spherically symmetric, static, black hole solutions of the theory defined by Eq. (\ref{eq:prepotential}). In order to do so we are going to use the so-called \emph{H-formalism}, developed in \cite{Galli:2011fq,Meessen:2011aa,Meessen:2012su}, based on the use of a new set of variables $H^M,~~ M=(\Lambda,\Lambda)$, that transform linearly under duality and reduce to harmonic functions on the transverse space $\mathbb{R}^{3}$ in the supersymmetric case.\footnote{It has been conjectured that this is also true in the extremal non-supersymmetric case  \cite{Meessen:2011aa,Meessen:2012su}.} This is the subject of the next section.

%%%%%%%%%%%%%%%%%%%%%%%%%%%%%%%%%%%%%%%%%%%%%%%%%%%%%%%%%%%%%%%%%%%%%%
%%%%%%%%%%%%%%%%%%%%%%%%%%%%%%%%%%%%%%%%%%%%%%%%%%%%%%%%%%%%%%%%%%%%%%
%%%%%%%%%%%%%%%%%%%%%%%%%%%%%%%%%%%%%%%%%%%%%%%%%%%%%%%%%%%%%%%%%%%%%%
%%%%%%%%%%%%%%%%%%%%%%%%%%%%%%%%%%%%%%%%%%%%%%%%%%%%%%%%%%%%%%%%%%%%%%

\subsection{A \emph{quantum} class of black holes}
\label{sec:blackholes}

%%%%%%%%%%%%%%%%%%%%%%%%%%%%%%%%%%%%%%%%%%%%%%%%%%%%%%%%%%%%%%%%%%%%%%
%%%%%%%%%%%%%%%%%%%%%%%%%%%%%%%%%%%%%%%%%%%%%%%%%%%%%%%%%%%%%%%%%%%%%%
%%%%%%%%%%%%%%%%%%%%%%%%%%%%%%%%%%%%%%%%%%%%%%%%%%%%%%%%%%%%%%%%%%%%%%
%%%%%%%%%%%%%%%%%%%%%%%%%%%%%%%%%%%%%%%%%%%%%%%%%%%%%%%%%%%%%%%%%%%%%%

The most general static, spherically symmetric space-time metric solution of an ungauged Supergravity is given by\footnote{The conformastatic coordinates $\left(t,\tau,\theta, \phi\right)$ cover the outer region of the event horizon when $\tau\in\left(-\infty, 0\right)$ and the inner region, between the Cauchy horizon and the physical singularity when $\tau\in\left(\tau_S, \infty\right)$, where $\tau_S\in\mathbb{R}^{+}$ is a model dependent number. The event horizon is located at $\tau \rightarrow -\infty$ and the Cauchy horizon at $\tau\rightarrow \infty$} \cite{Ferrara:1997tw,Meessen:2011aa}

\begin{equation}
\label{eq:generalbhmetric}
\begin{array}{rcl}
ds^{2}
& = &
e^{2U} dt^{2} - e^{-2U} \gamma_{\underline{m}\underline{n}}
dx^{\underline{m}}dx^{\underline{n}}\, ,  \\
& & \\
\gamma_{\underline{m}\underline{n}}
dx^{\underline{m}}dx^{\underline{n}}
& = &
{\displaystyle\frac{r_{0}^{4}}{\sinh^{4} r_{0}\tau}}d\tau^{2}
+
{\displaystyle\frac{r_{0} ^{2}}{\sinh^{2}r_{0}\tau}}d\Omega^{2}_{(2)}\, .\\
\end{array}
\end{equation}

\noindent
Using Eq. (\ref{eq:generalbhmetric}) and following the \emph{H-formalism}, we obtain that the equations of the theory are given by

\begin{equation}
\label{eq:Eqsofmotion}
\mathcal{E}_{P}=\tfrac{1}{2}\partial_{P}\partial_{M}\partial_{N}\log \mathsf{W}\,
\left[\dot{H}^{M}\dot{H}^{N} -\tfrac{1}{2}\mathcal{Q}^{M}\mathcal{Q}^{N}
\right]
+\partial_{P}\partial_{M}\log \mathsf{W}\, \ddot{H}^{M}
-\frac{d}{d\tau}\left(\frac{\partial \Lambda}{\partial \dot{H}^{P}}\right)
+\frac{\partial \Lambda}{\partial H^{P}}=0\, ,
\end{equation}

\noindent
together with the \emph{Hamiltonian constraint}

\begin{equation}
\label{eq:Hamiltonianconstraint}
\mathcal{H}\equiv
-\tfrac{1}{2}\partial_{M}\partial_{N}\log\mathsf{W}
\left(\dot{H}^{M}\dot{H}^{N}-\tfrac{1}{2}\mathcal{Q}^{M}\mathcal{Q}^{N} \right)
+\left(\frac{\dot{H}^{M}H_{M}}{ \mathsf{W}}\right)^{2}
-\left(\frac{\mathcal{Q}^{M}H_{M}}{ \mathsf{W}}\right)^{2}
-r_{0}^{2}=0\, ,
\end{equation}

\noindent
where

\begin{equation}
\Lambda \equiv \left(\frac{\dot{H}^{M}H_{M}}{ \mathsf{W}}\right)^{2}
+\left(\frac{\mathcal{Q}^{M}H_{M}}{ \mathsf{W}}\right)^{2}\, ,
\end{equation}

\noindent
and

\begin{equation}
\label{eq:W(H)}
\mathsf{W}(H) \equiv \mathcal{R}_{M}(H)H^{M} = e^{-2U}, ~~~\mathcal{R}+i\mathcal{I} =\mathcal{V}^M/X\, .
\end{equation}

\noindent
$\mathcal{V}^M$ is the covariantly holomorphic symplectic section of $\mathcal{N}=2$ Supergravity, and $X$ is a complex variable with the same K\"ahler weight as $\mathcal{V}^M$. $\mathcal{R}^M(\mathcal{I})$ stands for the real part $\left(\mathcal{R}^M\right)$ of $\mathcal{V}^M$ written as a function of the imaginary part $\mathcal{I}^M$, something that can always be done by solving the so-called \emph{stabilization equations}. $\mathsf{W}(H)$ is usually known in the literature as the \emph{Hesse potential}.

The theory is now expressed in terms of $2\left(n_{v}+1\right)$ variables $H^M$ and depends on $2\left(n_{v}+1\right)+1$ parameters: $2\left(n_{v}+1\right)$ charges $\mathcal{Q}^M$ and the non-extremality parameter $r_0$, from which one can reconstruct the solution in terms of the original fields of the theory (that is it, the space-time metric, scalars and vector fields).

For Eq. (\ref{eq:IIapert}), the general $\mathsf{W}(H)$ is an extremely involved function, and one cannot expect to solve in full generality the corresponding differential equations of motion, or even the associated algebraic equations of motion obtained by making use of the hyperbolic Ansatz for the $H^M$. Therefore, we are going to consider a particular truncation, which will give us the desired \emph{quantum} black holes

\begin{equation}
\label{eq:truncation}
H^0 = H_0 = H_i = 0,~~p^{0} = p_{0} = q_{i} = 0 \, .
\end{equation}

\noindent
Eq. (\ref{eq:truncation}) implies

\begin{equation}
\label{eq:hessian}
\mathsf{W}(H)=\alpha\left|\kappa^{0}_{ijk} H^i H^j H^k\right|^{2/3}\, ,
\end{equation}

\noindent
where $\alpha=\frac{\left( 3! c\right)^{1/3}}{2}$ must be positive in order to have a non-singular metric. Hence $c>0$ is a necessary condition in order to obtain  a regular solution and a consistent truncation. The corresponding \emph{black hole potential} reads

\begin{equation}
\label{eq:vbh}
V_{\rm bh} = \frac{\mathsf{W}(H)}{4}\partial_{ij}\log\mathsf{W}(H)\mathcal{Q}^{i}\mathcal{Q}^{j}\, ,
\end{equation}

\noindent
The scalar fields, purely imaginary, are given by

\begin{equation}
\label{eq:scalars}
z^i= i \left(3! c\right)^{1/3} \frac{H^i}{\left(\kappa^{0}_{ijk} H^i H^j H^k\right)^{1/3}} \, ,
\end{equation}

\noindent
and are subject to the following constraint, which ensures the regularity of the K\"ahler potential ($\mathcal{X}^0 = 1$ gauge)

\begin{equation}
\label{eq:kahlercondition}
\kappa^{0}_{ijk}\Im{\rm m}z^{i}\Im{\rm m}z^{j}\Im{\rm m}z^{k} > \frac{3c}{2}\, .
\end{equation}

\noindent
Substituting Eq. (\ref{eq:scalars}) into Eq. (\ref{eq:kahlercondition}), we obtain

\begin{equation}
\label{eq:kahlerconditionII}
c>\frac{c}{4}\ \, ,
\end{equation}

\noindent
which is an identity (assuming $c>0$) and therefore imposes no constraints on the scalars. This phenomenon can be traced back to the fact that the the K\"ahler potential is constant when evaluated on the solution, and given by

\begin{equation}
\label{eq:kahlerpotential}
e^{-\mathcal{K}}=6c\ \, ,
\end{equation}

\noindent
which is well defined, again, if $c>0$. Since the volume of the C.Y. manifold is proportional to $e^{-\mathcal{K}}$, Eq. (\ref{eq:kahlerpotential}) implies that such volume remains constant and, in particular, that the limit $\Im{\rm m}z^{i}\rightarrow \infty$ does not imply a large volume limit of the compactification C.Y. manifold, a remarkable fact that can be seen as a purely quantum characteristic of our solution\footnote{Notice that in order to consistently discard the non-perturbative terms in Eq. (\ref{eq:IIaprepotential}) we only need to take the limit $\Im{\rm m}z^{i}\rightarrow \infty$. Therefore, the behavior of the C.Y. volume in such limit plays no role.}. Notice that it is also possible to obtain the classical limit $\Im{\rm m}z^{i}\gg1$ taking $c\gg1$, that is, choosing a Calabi-Yau manifold with large enough $c$. In this case we would have also a truly large volume limit.

We have seen that, in order to obtain a consistent truncation, a necessary condition is $c>0$, which implies that $\mathsf{W}\left(H\right)$ is well defined. We can go even further and argue that this is a sufficient condition by studying the equations of motion $\mathcal{E}_{P}$: 

A consistent truncation requires that the equation of motion of the truncated field is identically solved for the \emph{truncation value} of the field. First, notice that the set of solutions of Eqs. (\ref{eq:Eqsofmotion}) and (\ref{eq:Hamiltonianconstraint}), taking into account (\ref{eq:truncation}), is non-empty, since there is a \emph{model-independent} solution, given by

\begin{equation}
\label{eq:universalsusy}
H^{i} = a^{i}-\frac{p^{i}}{\sqrt{2}}\tau,~~~~r_0=0\, ,
\end{equation}

\noindent
which corresponds to a supersymmetric black hole. However, the equations of motion $\mathcal{E}_{P}$ don't know about supersymmetry: it is system of differential equations whose solution can be written as

\begin{equation}
\label{eq:universalsolution}
H^{M} = H^M\left(a,b\right)\, ,
\end{equation}

\noindent
where we have made explicit the dependence in $2n_{v}+2$ integration constants. When the solution (\ref{eq:universalsolution}) is plugged into (\ref{eq:Hamiltonianconstraint}) is when we impose, through $r_0$, a particular condition about the extremality of the black hole. If $r_0 = 0$ the integration constants are fixed such as the solution is extremal. In general there is not a unique way of doing it, one of the possibilities being always the supersymmetric one. Therefore, given that for our particular truncation the supersymmetric solution always exists, we can expect the existence also of the corresponding solution (\ref{eq:universalsolution}) of the equations of motion, from which the supersymmetric solution may be obtained through a particular choice of the integration constants that make (\ref{eq:universalsolution}) fulfilling (\ref{eq:Hamiltonianconstraint}) for $r_{0} = 0$. 

\noindent
We conclude, hence, that

\begin{equation}
\label{eq:consistentproof}
\left\{ H^{P} = 0,\mathcal{Q}^P=0 \right\} ~~\Rightarrow \mathcal{E}_{P}=0\, ,
\end{equation}

\noindent
and therefore the truncation of as many $H$'s as we want, together with the correspondet $\mathcal{Q}$'s, is consistent as long as $\mathsf{W}\left(H\right)$ remains well defined, something that in our case is assured if $c>0$. From Eq. (\ref{eq:IIaprepotential}) it can be checked that the case $c=0$, that is $h^{1,1}=h^{2,1}$,  can be cured by non-perturbative effects. 

%The existence of regular non-extremal black holes is expected but has to be checked in a model-dependent way.

It is easy to see that the truncation is not consistent in the classical limit, and therefore, we can conclude that the corresponding solutions are \emph{genuinely} quantum solutions, which only exist when perturbative quantum effects are incorporated into the action. 

Hence, we can conclude that if we require our theory to contain regular \emph{quantum} black holes there is a topological restriction on the Calabi-Yau manifolds that we can choose to compactify Type-IIA String Theory. The condition can be expressed as

\begin{equation}
\label{eq:conditionc}
c>0 ~~ \Rightarrow ~~ h^{11}>h^{21} \, .
\end{equation}

\noindent
Eq. (\ref{eq:conditionc}) is a stringent condition on the compactification C.Y. manifolds, in particular for small $h^{11}$. In fact, for small enough $h^{11}$ it could be even possible that no Calabi--Yau manifold existed such Eq. (\ref{eq:conditionc}) is fulfilled. We will investigate this issue for $h^{11} =  3$, explicitly constructing the corresponding C.Y. manifolds and finding also particular \emph{quantum} black hole solutions, in the next section\footnote{It is known in the literature the existence of the so called \emph{rigid} C.Y. manifolds \cite{Candelas:1985en,Strominger:1985it,Candelas:1993nd}, which obey $h^{11}>0, h^{21}=0$, being therefore admissible compactification spaces. However, in order to have a tractable theory, we need a small enough $h^{11}$, yet not too small to yield a trivial theory. The choice $h^{11}=3$ fulfills both conditions.}.

%%%%%%%%%%%%%%%%%%%%%%%%%%%%%%%%%%%%%%%%%%%%%%%%%%%%%%%%%%%%%%%%%%%%%%
%%%%%%%%%%%%%%%%%%%%%%%%%%%%%%%%%%%%%%%%%%%%%%%%%%%%%%%%%%%%%%%%%%%%%%
%%%%%%%%%%%%%%%%%%%%%%%%%%%%%%%%%%%%%%%%%%%%%%%%%%%%%%%%%%%%%%%%%%%%%%
%%%%%%%%%%%%%%%%%%%%%%%%%%%%%%%%%%%%%%%%%%%%%%%%%%%%%%%%%%%%%%%%%%%%%%

\section{New Calabi--Yau manifolds}
\label{sec:NCY}

%%%%%%%%%%%%%%%%%%%%%%%%%%%%%%%%%%%%%%%%%%%%%%%%%%%%%%%%%%%%%%%%%%%%%%
%%%%%%%%%%%%%%%%%%%%%%%%%%%%%%%%%%%%%%%%%%%%%%%%%%%%%%%%%%%%%%%%%%%%%%
%%%%%%%%%%%%%%%%%%%%%%%%%%%%%%%%%%%%%%%%%%%%%%%%%%%%%%%%%%%%%%%%%%%%%%
%%%%%%%%%%%%%%%%%%%%%%%%%%%%%%%%%%%%%%%%%%%%%%%%%%%%%%%%%%%%%%%%%%%%%%

In this section we will present the construction of new Calabi--Yau manifolds which
satisfy $h^{1,1} = 3$ and $h^{2,1} < 3$, as required for the truncation
presented in the previous section.

Calabi--Yau threefolds with both Hodge numbers small are relatively rare; two
large and useful databases are the complete intersections in products of projective
spaces (CICY's) \cite{Candelas:1987kf}, and hypersurfaces in toric fourfolds
\cite{Batyrev:1994hm,Kreuzer:2000xy}, but the
manifolds in these lists all satisfy the inequality $h^{1,1} + h^{2,1} > 21$.  Smaller Hodge
numbers can be found by taking quotients by groups which have a free
holomorphic action on one of these manifolds
(see, e.g., \cite{Candelas:2008wb,Braun:2010vc,Davies:2011fr} and
references therein), but none of the known spaces constructed this way satisfy our
requirements.

Our technique here will be to begin with known manifolds with
$h^{1,1} < 3$, $h^{2,1} < 4$, and a non-trivial fundamental group, and find
hyperconifold transitions \cite{Davies:2009ub,Davies:2011is} to new manifolds
with the required Hodge numbers.  Briefly, these transitions occur because a
generically-free group action on a Calabi--Yau will develop fixed points on
certain codimension-one loci in the moduli space.  The fixed points are
necessarily singular, and typically nodes \cite{Davies:2009ub}, so the
quotient space develops a point-like singularity which is a quotient of the
conifold --- a \emph{hyperconifold}.  These singularities can be resolved to
give a new smooth Calabi--Yau.  If the subgroup which develops a fixed point
is $\IZ_N$, then the change in Hodge numbers for one of these transitions
is $\d\hodgenos = (N-1, -1)$.

Interestingly, there are examples which one
might na\"ively believe would lead to manifolds with $\hodgenos = (3,0)$ and
$\hodgenos = (3,2)$, but none of these work out;\footnote{In each case, the
spaces have unavoidable symmetries which make it impossible to create just
a single hyperconifold singularity.  Resolving the extra singularities pushes
$h^{1,1}$ higher.} instead, we have two examples with $\hodgenos = (3,1)$,
but with different intersection forms and K\"ahler cones.

%%%%%%%%%%%%%%%%%%%%%%%%%%%%%%%%%%%%%%%%%%%%%%%%%%%%%%%%%%%%%%%%%%%%%%
%%%%%%%%%%%%%%%%%%%%%%%%%%%%%%%%%%%%%%%%%%%%%%%%%%%%%%%%%%%%%%%%%%%%%%

\subsection{\texorpdfstring{$\hodgenos = (3,1)$}{(h11, h21) = (3,1)} and diagonal intersection form}

%%%%%%%%%%%%%%%%%%%%%%%%%%%%%%%%%%%%%%%%%%%%%%%%%%%%%%%%%%%%%%%%%%%%%%
%%%%%%%%%%%%%%%%%%%%%%%%%%%%%%%%%%%%%%%%%%%%%%%%%%%%%%%%%%%%%%%%%%%%%%

For the first example, we start with a manifold $X^{1,3}$, where the superscripts
are the Hodge numbers $\hodgenos$, and fundamental group
$\IZ_5{\times}\IZ_2{\times}\IZ_2 \cong \IZ_{10}{\times}\IZ_2$.  It was first
discovered in \cite{Candelas:2008wb}, and we briefly review the construction here.
The manifold is obtained as a free quotient of a CICY $X^{5,45}$ that is given
by the vanishing of two multilinear polynomials in a product of five $\IP^1$'s; the
configuration matrix \cite{Candelas:1987kf} is
\begin{equation*}
    \cicy{\IP^1 \\ \IP^1 \\ \IP^1 \\ \IP^1 \\ \IP^1}
        {1 & 1 \\
         1 & 1 \\
         1 & 1 \\
         1 & 1 \\
         1 & 1}
\end{equation*}

Let us call the two polynomials $p_1, p_2$, and take homogeneous coordinates
$t_{i,a}$ on the ambient space, where $i = 0,1,2,3,4$ is understood mod 5, and
$a = 0,1$ is understood mod 2.  Then the action of the quotient group is generated by
\begin{align*}
    g_{10} ~&:~ t_{i,a} \to t_{i+1, a+1} ~~~~;~~ p_1 \leftrightarrow p_2 ~, \\
    g_2 ~&:~ t_{i,a} \to (-1)^{a} t_{i,a} ~~;~~ p_1 \to p_1 ~,~ p_2 \to -p_2 ~.
\end{align*}
Note that these commute only up to projective equivalence, but this is sufficient.
To define polynomials which transform appropriately, we start with the following
quantities:
\begin{equation*}
    m_{abcde} = \sum_{i=0}^4 t_{i,a} t_{i+1,b} t_{i+2,c} t_{i+3,d} t_{i+4,e} ~.
\end{equation*}
Then it is easily checked that the following are the most general polynomials which
transform correctly:
\begin{align*}
    p_1 &= \frac{A_0}{5} m_{00000} + A_1 m_{00011} + A_2 m_{00101} + A_3 m_{01111} ~,\\
    p_2 &= \frac{A_0}{5} m_{11111} + A_1 m_{11100} + A_2 m_{11010} + A_3 m_{10000} ~,
\end{align*}
where the $A_\a$ are arbitrary complex constants.  For generic values of the coefficients,
these polynomials define a smooth manifold on which the group $\IZ_{10}{\times}\IZ_2$
acts freely; in this way we find a smooth quotient family
$X^{1,3} = X^{5,45}/\IZ_{10}{\times}\IZ_2$.

We now need to specialise to a sub-family of $X^{5,45}$ which \emph{does} have fixed
points of the group generator $g_2$.  Specifically, consider the point given by
\begin{equation*}
    t_{0,1} = t_{1,1} = t_{2,1} = t_{3,0} = t_{4,0} = 0,
\end{equation*}
which is fixed by the action of $g_2$.  Substituting the above into the polynomials
gives the values $p_1 = A_1,~ p_2 = 0$, so if we set $A_1 = 0$, $X^{5,45}$ will contain this
point.  The argument of \cite{Davies:2009ub} guarantees that it will be a
singularity, and one can check that for general values of the other coefficients, it is a
node, so the quotient space $X^{1,3}$ develops a $\IZ_2$-hyperconifold singularity.  In
fact, there are nine other points related to the above by the action of the other group
generator $g_{10}$, so the covering space $X^{5,45}$ actually has ten nodes.  Since
these are all identified by the group action, $X^{1,3}$ develops only a single
$\IZ_2$-hyperconifold.  This can be resolved by a single blow-up, and we obtain a new
manifold with Hodge numbers $\hodgenos = (2,2)$, and fundamental group $\IZ_{10}$.

To get all the way to $X^{3,1}$, we need to go through another $\IZ_2$-hyperconifold
transition.  If we also set $A_2 = 0$, then $X^{5,45}$ also passes through another
fixed point of $g_2$, given by
\begin{equation*}
    t_{0,1} = t_{1,1} = t_{2,0} = t_{3,1} = t_{4,0} = 0 ~,
\end{equation*}
as well as the nine points related to this by the action of $g_{10}$.

It can be checked that when $A_1 = A_2 = 0$, $X^{5,45}$ has exactly twenty
nodes, at the points described above, and is smooth elsewhere.  Therefore $X^{1,3}$
has precisely two $\IZ_2$-hyperconifold singularities, which we can resolve
independently to obtain a new smooth Calabi--Yau manifold $X^{3,1}$.

%%%%%%%%%%%%%%%%%%%%%%%%%%%%%%%%%%%%%%%%%%%%%%%%%%%%%%%%%%%%%%%%%%%%%%

\subsubsection{The intersection form and K\"ahler cone}
\label{subKc1}
%%%%%%%%%%%%%%%%%%%%%%%%%%%%%%%%%%%%%%%%%%%%%%%%%%%%%%%%%%%%%%%%%%%%%%

To find the Supergravity theory coming from compactification on $X^{3,1}$, we
need to calculate its triple intersection form, and for this we need a basis for
$H^2(X^{3,1}, \IZ)$ (throughout, we will implicitly talk about only the torsion-free
part of the cohomology).  There is a natural basis which consists of one divisor
class inherited from $X^{1,3}$, and the two exceptional divisor classes coming
from the two blow-ups.

First, let us find an integral generator of $H^2(X^{1,3}, \IZ)$.  On the covering
space $X^{5,45}$, let $H_i$ be the divisor class given by the pullback of the
hyperplane class from the $i^{\mathrm{th}}$ $\IP^1$.  Then the invariant divisor
classes are multiples of $H \equiv H_0 + H_1 + H_2 + H_3 + H_4$.  However,
$H$ itself, although an invariant \emph{class}, does not have an invariant
\emph{representative}.  The class $2H$ does, however; an example of an
invariant divisor in $2H$ is the surface given by the vanishing of
\begin{equation*}
    f = (m_{00011})^2 + (m_{11100})^2 + (m_{00101})^2 + (m_{11010})^2 ~.
\end{equation*}
This is a particularly convenient choice, as it gives a smooth divisor even on
the singular family of threefolds given by $A_1 = A_2 = 0$, which misses the
singular points.

Let $D_1$ be the divisor given by setting $f = 0$ and then taking the quotient,
and let $D_2$ and $D_3$ be the two exceptional divisors.  Then, since $f \neq 0$
on the fixed points of the group action, we immediately see that $D_1, D_2$, and
$D_3$ are all disjoint, and the only intersection numbers which might be
non-zero are $D_1^3$, $D_2^3$, and $D_3^3$.

For $D_2^3$ and $D_3^3$, we can use an easy general argument.  For any
smooth surface $S$ in a Calabi--Yau threefold, the adjunction formula gives
$S\rest{S} \sim K_S$, where $K_S$ is the canonical divisor class.  The
triple intersection number $S^3$ is therefore equal to $K_S^2$.  Each of
$D_2$ and $D_3$ is isomorphic to $\IP^1{\times}\IP^1$, so we find
$D_2^3 = D_3^3 = 8$.

To calculate $D_1^3$, we note that $D_1$ descends from the divisor
class $2H$ on $X^{5,45}$.  Since this is embedded in a product of projective
spaces, we can calculate intersection numbers purely from degrees; it is easy
to check that on $X^{5,45}$, $(2H)^3 = 960$.  We divide by a freely-acting
group of order twenty, so on the quotient space we find
$D_1^3 = \frac{960}{20} = 48$.

To summarise, the non-vanishing triple intersection numbers of $X^{3,1}$, in
the basis ${D_1, D_2, D_3}$, are
\begin{equation*}
    \k^0_{111} = 48 ~,~~ \k^0_{222} = \k^0_{333} = 8 ~.
\end{equation*}

We can also say something about the K\"ahler cone.  Certainly $D_1$ is
positive everywhere except on the exceptional divisors, where it is trivial.  On
the other hand, each exceptional divisor $D$ contains curves $C$ for which
$D\cdot C = -1$.  From this information, we can glean that the K\"ahler cone is
some sub-cone of $t_1 > 0, t_2 < 0, t_3 < 0$, and certainly includes the region
where $t_1$ is much larger than $|t_2|$ and $|t_3|$.

%%%%%%%%%%%%%%%%%%%%%%%%%%%%%%%%%%%%%%%%%%%%%%%%%%%%%%%%%%%%%%%%%%%%%%
%%%%%%%%%%%%%%%%%%%%%%%%%%%%%%%%%%%%%%%%%%%%%%%%%%%%%%%%%%%%%%%%%%%%%%

\subsection{\texorpdfstring{$\hodgenos = (3,1)$}{(h11, h21) = (3,1)} and non-diagonal intersection form}

%%%%%%%%%%%%%%%%%%%%%%%%%%%%%%%%%%%%%%%%%%%%%%%%%%%%%%%%%%%%%%%%%%%%%%
%%%%%%%%%%%%%%%%%%%%%%%%%%%%%%%%%%%%%%%%%%%%%%%%%%%%%%%%%%%%%%%%%%%%%%

For our second example, we will again start with a free quotient of a CICY
manifold, with configuration matrix
\begin{eqnarray*}
    &\cicy{\IP^1 \\ \IP^1 \\\IP^1 \\\IP^1 \\\IP^1 \\\IP^1 \\\IP^1}
    {
    0 & 0 & 1 & 1 \\
    1 & 0 & 1 & 0 \\
    1 & 0 & 1 & 0 \\
    1 & 0 & 1 & 0 \\
    0 & 1 & 0 & 1 \\
    0 & 1 & 0 & 1 \\
    0 & 1 & 0 & 1}
    \\[.2ex]
    &\hspace{1em} p~~\,q~\,~r_1~r_2
\end{eqnarray*}
where the labels on the columns denote the respective polynomials.  This
manifold has Euler number zero, and a series of splittings and contractions
(explained in \cite{Candelas:1987kf,Candelas:2008wb,Candelas:2010ve})
establishes that it is in fact isomorphic to the `split bicubic' or Schoen manifold,
with Hodge numbers $\hodgenos = (19,19)$.

Let us take homogeneous coordinates $\s_a$ on the first $\IP^1$,
$s_{i,a}$ on the next three, and $t_{i,a}$ on the last
three, where $i = 0,1,2~,~\mathrm{and}~ a = 0,1$ are understood mod 3
and mod 2 respectively.  The quotient group of interest is the dicyclic group
$\Dic_3 \cong \IZ_3 \rtimes \IZ_4$, which is the only non-trivial semi-direct
product of $\IZ_3$ and $\IZ_4$.  It is generated by two elements $g_3$ and
$g_4$, of orders given by their subscripts, with the relation
$g_4 g_3 g_4^{-1} = g_3^2$, and acts on the ambient space and
polynomials as follows:
\begin{align*}
    g_3 ~&:~ \s_a \to \s_a ~,~ s_{i,a} \to s_{i+1, a} ~,~ t_{i,a} \to t_{i+1,a}
        ~;~ \text{all polynomials invariant}~,\\[1ex]
    g_4 ~&:~ \s_a \to (-1)^a \s_a ~,~ s_{i,a} \to (-1)^{a+1} t_{-i,a} ~,~ t_{i,a} \to s_{-i,a}
        ~;~ p \to -q ~,~ q\to p ~,~ r_1 \leftrightarrow r_2 ~.
\end{align*}

In order to write down polynomials which transform appropriately, let us
first define the $g_3$-invariant quantities
\begin{equation*}
    m_{abc} = \sum_i s_{i,a} s_{i+1,b} s_{i+2,c} ~~,~ n_{abc} = \sum_i t_{i,a} t_{i+1,b} t_{i+2,c} ~.
\end{equation*}
Then by choice of coordinates (consistent with the above action), we can
take the polynomials to be
\begin{align*}
    p = \frac 13 m_{000} + m_{011} \quad,\quad q = \frac 13 n_{000} + n_{011} \qquad,\\[1ex]
    r_1 = (a_0 m_{001} + \frac 13 a_1 m_{111})\s_0 + (a_0 m_{001} + \frac 13 a_2 m_{111})\s_1 ~~,\\[1ex]
    r_2 = (a_0 n_{001} + \frac 13 a_1 n_{111})\s_0 - (a_0 n_{001} + \frac 13 a_2 n_{111})\s_1 \quad,
\end{align*}
where $a_0, a_1, a_2$ are arbitrary complex coefficients, defined only up
to overall scale.  It can be checked that for generic values of these coefficients,
the corresponding manifold is smooth, and the group acts on it without fixed
points.  We therefore obtain a smooth quotient manifold $X^{2,2}$, where the
value $h^{2,1} = 2$ corresponds to the two free coefficients (once we factor out
overall scale) in the above polynomials.\footnote{Counting independent
coefficients does not always give the value of $h^{2,1}$, but in this case it does;
perhaps the most direct way to obtain this is to notice that the manifold is
obtained via a conifold transition on a codimension two locus in the moduli space
of a manifold $X^{1,4}$, which was described at length in \cite{Braun:2009qy}.}

We will now show that there is a $\IZ_2$-hyperconifold transition from $X^{2,2}$
to a manifold with $\hodgenos = (3,1)$.  To do this, we need to arrange for the
unique order-two element, $g_4^2$, to develop a fixed point.  Consider the point
in the ambient space given by
\begin{equation*}
    \frac{\s_1}{\s_0} = -1 ~,~ s_{0,1} = s_{1,1} = s_{2,0} = t_{0,0} = t_{1,0} = t_{2,0} = 0~.
\end{equation*}
This is fixed by $g_4^2$, but the other elements of the group permute this and
five other $g_4^2$-fixed points.  If we evaluate the polynomials at the point above,
we find
\begin{equation*}
    p = q = r_1 = 0 ~,~ r_2 = a_1 + a_2 ~,
\end{equation*}
and their values at the other five fixed points are related by the group action to
the ones above.  So if $a_1 + a_2 = 0$, the Calabi--Yau will intersect these
fixed points.  By expanding the polynomials around any one of these points, we
find that it has a node at each of them, so on the quotient space, we obtain a
single $\IZ_2$-hyperconifold singularity.  Resolving this takes us to a new
smooth manifold $Y^{3,1}$ (we use the letter $Y$ to distinguish this from the
other $(3,1)$ manifold we constructed).  Its fundamental group is
$\Dic_3/\langle g_4^2\rangle \cong S_3$, the symmetric group on three letters
(the behaviour of fundamental groups under hyperconifold transitions such as
this one is described in \cite{Davies:2011fr}).

%%%%%%%%%%%%%%%%%%%%%%%%%%%%%%%%%%%%%%%%%%%%%%%%%%%%%%%%%%%%%%%%%%%%%%

\subsubsection{The intersection form and K\"ahler cone}
\label{subKc2}
%%%%%%%%%%%%%%%%%%%%%%%%%%%%%%%%%%%%%%%%%%%%%%%%%%%%%%%%%%%%%%%%%%%%%%

To calculate the intersection form of $Y^{3,1}$, we start with $X^{2,2}$
and its covering space $X^{19,19}$.  Part of $H^{1,1}(X^{19,19}, \IZ)$ is
generated by the pullbacks of the hyperplane classes of the $\IP^1$ spaces.
We will denote these by ${H_0, H_1, \ldots, H_6}$.  Looking at the group action,
we can see that there are exactly two invariant divisor classes constructed from
these: $H_0$ and $H_1 + H_2 + \ldots + H_6$.  In contrast to the last example,
each of these actually contains an invariant representative, and we get a basis
$\{D_1, D_2\}$ for $H^{1,1}(X^{2,2},\IZ)$ by simply taking the two invariant
classes above and quotienting.

On the covering space, we can calculate intersection numbers simply by
counting degrees, and we find that
\begin{equation*}
    H_0(H_1 + H_2 + \ldots + H_6)^2 = 72 ~,~ (H_1 + H_2 + \ldots + H_6)^3 = 216 ~,
\end{equation*}
and all others vanish.  Dividing by the order of the group, we see that on the
quotient space
\begin{equation*}
    D_1D_2^2 = 6 ~,~ D_2^3 = 18 ~,
\end{equation*}
and the other triple intersections are zero.

Finally, we perform the transition to $Y^{3,1}$; denote the class of the exceptional
divisor by $D_3$.  It is easy enough to check that $D_1$ and $D_2$ have
representatives which miss the singularity, so their pullbacks to $Y^{3,1}$ are
disjoint from the exceptional divisor, and we get
$D_1\cdot D_3 = D_2\cdot D_3 = 0$.  Once again, the exceptional divisor is
isomorphic to $\IP^1{\times}\IP^1$, so by the argument of the last section,
$D_3^3 = 8$.

Summarising, the non-zero intersection numbers on $Y^{3,1}$ are
\begin{equation*}
    \k^0_{122} = 6 ~,~ \k^0_{222} = 18 ~,~ \k^0_{333} = 8 ~.
\end{equation*}

By similar reasoning to the last case, we can say that the K\"ahler cone is some
sub-cone of $t_1 > 0, t_2 > 0, t_3 < 0$, and includes the region
where $|t_3|$ is sufficiently small compared to $t_1$ and~$t_2$.

%%%%%%%%%%%%%%%%%%%%%%%%%%%%%%%%%%%%%%%%%%%%%%%%%%%%%%%%%%%%%%%%%%%%%%
%%%%%%%%%%%%%%%%%%%%%%%%%%%%%%%%%%%%%%%%%%%%%%%%%%%%%%%%%%%%%%%%%%%%%%
%%%%%%%%%%%%%%%%%%%%%%%%%%%%%%%%%%%%%%%%%%%%%%%%%%%%%%%%%%%%%%%%%%%%%%
%%%%%%%%%%%%%%%%%%%%%%%%%%%%%%%%%%%%%%%%%%%%%%%%%%%%%%%%%%%%%%%%%%%%%%

\section{Quantum black hole solutions with \texorpdfstring{$h^{11}=3$}{h11 = 3}}
\label{sec:BH3}

%%%%%%%%%%%%%%%%%%%%%%%%%%%%%%%%%%%%%%%%%%%%%%%%%%%%%%%%%%%%%%%%%%%%%%
%%%%%%%%%%%%%%%%%%%%%%%%%%%%%%%%%%%%%%%%%%%%%%%%%%%%%%%%%%%%%%%%%%%%%%
%%%%%%%%%%%%%%%%%%%%%%%%%%%%%%%%%%%%%%%%%%%%%%%%%%%%%%%%%%%%%%%%%%%%%%
%%%%%%%%%%%%%%%%%%%%%%%%%%%%%%%%%%%%%%%%%%%%%%%%%%%%%%%%%%%%%%%%%%%%%%

In section \ref{sec:generaltheory} we have presented a particular truncation of the equations of motion of $\mathcal{N}=2, d=4$ ungauged Supergravity in a static, spherically symmetric background, which turned out to be consistent only for positive values of the quantum perturbative coefficient $c$ (\ref{eq:conditionc}).  In the next two sections we are going to explicitly construct regular non-extremal (and therefore non-supersymmetric) black hole solutions to the truncated theory. In particular, we will start studying the cases where the C.Y. manifold is of the type constructed in section  \ref{sec:NCY}, to wit:

\begin{equation*}
    X^{3,1}\Rightarrow\k^0_{111} = 48 ~,~ \k^0_{222} =\k^0_{333} = 8\, ,
\end{equation*}

\begin{equation*}
    Y^{3,1}\Rightarrow\k^0_{122} = 6 ~,~ \k^0_{222} = 18 ~,~ \k^0_{333} = 8\, .
\end{equation*}

\noindent
For these two sets of intersection numbers, Eq. (\ref{eq:hessian}) becomes, respectively

\begin{equation}
\mathsf{W}(H)=\alpha\left|48\left(H^1\right)^{3}+8 \left[\left(H^2\right)^{3}+ \left(H^3\right)^{3}\right]\right|^{2/3}\, ,
\end{equation}

\begin{equation}
\mathsf{W}(H)=\alpha\left|18 \left(H^2\right)^2\left[{H^1}+{H^2}\right]+8\left(H^3\right)^{3}\right|^{2/3}\, .
\end{equation}

\noindent
For simplicity we take $H^1=s_2 H^2=s_3 H^3 \equiv H$ ($s_{2,3}=\pm 1$), $p^1=p^2=p^3\equiv p$. For this particular configuration we find a non-extremal solution for each set of intersection numbers given by

\begin{equation}
H=a \cosh (r_0 \tau) +\frac{b}{r_0}\sinh(r_0 \tau),~~~~ b=s_b \sqrt{r_0^2 a^2+\frac{p^2}{2}}\, ,
\end{equation}

\noindent
where, now and henceforth, $s_b=\pm 1$. The scalars, which turn out to be constant, read

\begin{equation}
z^1=i(3! c)^{1/3} \lambda^{-1/3}=s_{2,3} z^{2,3}\, ,
\end{equation}

\noindent
where

\begin{align}
&\lambda = \left[48+8(s_2+s_3)\right]~ ~\text{for}~~X^{3,1},\\ \notag
&\lambda= \left[18+18s_2+8s_3\right]~~\text{for}~~Y^{3,1}\, .
\end{align}

\noindent
Since the scalars are constant and don't depend on the charges, we cannot perform the $\Im{\rm m}z^{i}\rightarrow\infty$ limit that fully suppress the non-perturbative corrections. Still, the exponent in Eq. (\ref{eq:IIanonpert}) is, in both cases, of order 

\begin{equation}
\label{eq:ordernonpert}
  2\pi i d_{i} z^{i} \sim -\frac{1}{3}\sum^{3}_{i=1}d_{i},~~~d_{i}\ge 1 \, ,
\end{equation}

\noindent
and therefore we find small non-perturbative corrections, in particular one order smaller than the perturbative part of the prepotential $\mathcal{F}_{Pert} \sim 10\cdot\mathcal{F}_{Non-Pert} $ . The solution lies inside the K\"ahler cone when 

\begin{equation}
\label{eq:cond1}
    s_2=s_3=-1, ~~ \text{for}~ X^{3,1}\, ,
\end{equation}

\begin{equation}
\label{eq:cond2}
    s_2=-s_3=1, ~~ \text{for}~ Y^{3,1}\, .
\end{equation}

\noindent
This can be verified by explicitly checking the positive-definiteness of the K\"ahler metric 
\begin{equation}
\mathcal{G}_{ij^*}=\partial_i\partial_{j^*}\mathcal{K}
\end{equation}
evaluated on the solution. It turns out that the only sets of $\{s_2,s_3\}$ which give rise to positive-definite K\"ahler metrics (and, as a consequence, to solutions lying inside the K\"ahler cone) are the ones shown above. These conditions on the signs of the scalar fields are in full agreement with those obtained in subsections (\ref{subKc2}) and (\ref{subKc1}), since $\Im m z^i=t^i$ \cite{Candelas:1990rm}. 

Imposing asymptotic flatness, the constant $a$ gets fixed to

\begin{equation}
a=-s_b \frac{\Im m z^1}{\sqrt{3c}}\, .
\end{equation}

\noindent
It is now easy to compute the mass and the entropy of the outer/inner horizon

\begin{equation}
\label{eq:mass48}
M=r_0\sqrt{1+\frac{3cp^2}{2r_0^2(\Im{\rm m} z^1)^2}}\, ,
\end{equation}

\begin{equation}
\label{eq:entropy48}
S_{\pm} = r_0^2\pi\left(\sqrt{1+\frac{3cp^2}{2r_0^2 (\Im{\rm m} z^1)^2}}\pm 1  \right)^{2}\, .
\end{equation}
This implies that the product of both entropies only depends on the charge

\begin{equation}
\label{eq:entropy48product}
S_{+} S_{-} = \frac{\pi^2\alpha^2}{4}p^{4}\lambda^{4/3}\, .
\end{equation}

It is worth stressing that the Ansatz $H^{i}=a^{i}+b^{i}\tau$ in the extremal ($r_0 = 0$) case was successfully used to obtain solutions with constant scalars but different critical points, in some cases particularly involved. However, presumably due to the complexity of the calculations, we have not been able to find a solution with non-constant scalars for any of the two models analyzed in this section. This may suggest also a more stabilized behavior for the scalars in the presence of perturbative quantum corrections.

\section{Quantum corrected \texorpdfstring{$STU$}{STU} model}
\label{sec:STUmodel}

%%%%%%%%%%%%%%%%%%%%%%%%%%%%%%%%%%%%%%%%%%%%%%%%%%%%%%%%%%%%%%%%%%%%%%
%%%%%%%%%%%%%%%%%%%%%%%%%%%%%%%%%%%%%%%%%%%%%%%%%%%%%%%%%%%%%%%%%%%%%%
%%%%%%%%%%%%%%%%%%%%%%%%%%%%%%%%%%%%%%%%%%%%%%%%%%%%%%%%%%%%%%%%%%%%%%
%%%%%%%%%%%%%%%%%%%%%%%%%%%%%%%%%%%%%%%%%%%%%%%%%%%%%%%%%%%%%%%%%%%%%%

In this section we consider a very special case, the so-called $STU$ model, in the presence of perturbative quantum corrections, obtaining the first non-extremal solution with non-constant scalars. In order to do so, we set $n_v=3,~ \kappa^{0}_{123}=1$. From (\ref{eq:hessian}) we obtain\footnote{We have to stress that we haven't been able to construct an explicit C.Y. manifold with $\kappa^{0}_{123}=1$ and $h^{21}<3$.}

\begin{equation}
\label{eq:truncationSTU}
\mathsf{W}(H)=\alpha\left|H^1 H^2 H^3\right|^{2/3}\ \ ,
\end{equation}

\noindent
where $\alpha=3 c^{1/3}$. The scalar fields are given by

\begin{equation}
\label{eq:scalarsSTU}
z^i= i c^{1/3} \frac{H^i}{\left(H^1 H^2 H^3\right)^{1/3}} \ \ ,
\end{equation}

\noindent
The $\tau$-dependence of the $H^M$ can be found by solving Eqs. (\ref{eq:Eqsofmotion}) and (\ref{eq:Hamiltonianconstraint}), and the solution is given by 

\begin{equation}
\label{eq:solutionHSTU}
H^i = a^i \cosh \left(r_0\tau\right) + \frac{b^i}{r_0} \sinh \left(r_0\tau\right), ~~~~ b^i=s^i_b\sqrt{r^2_0 (a^i)^2+\frac{(p^i)^2}{2}}\ \ .
\end{equation}

\noindent
The three constants $a^i$ can be fixed relating them to the value of the scalars at infinity and imposing asymptotic flatness. We have, hence, four conditions for three parameters and therefore one would expect a relation among the $\Im{\rm m}z^i_{\infty}$, leaving $c$ undetermined. However, the explicit calculation shows that the fourth relation is compatible with the others, and therefore no extra constraint is necessary. The $a^{i}$ are given by

\begin{equation}
\label{eq:askSTU}
a^i=-s^i_b\frac{\Im{\rm m}z_{\infty}^i}{\sqrt{3 c}}\ \ .
\end{equation}

\noindent
The mass and the entropy, in turn, read

\begin{equation}
\label{eq:massSTU}
M=\frac{r_0}{3}\sum_i \sqrt{1+\frac{3c(p^i)^2}{2r_0^2(\Im{\rm m} z_{\infty}^i)^2}}\ \ ,
\end{equation}

\begin{equation}
\label{eq:entropySTU}
S_{\pm} = r_0^2\pi\prod_i \left(\sqrt{1+\frac{3c(p^i)^2}{2r_0^2 (\Im{\rm m} z_{\infty}^i)^2}}\pm 1 \right)^{2/3}\ \ ,
\end{equation}

\noindent
and therefore the product of the inner and outer entropy only depends on the charges

\begin{equation}
\label{eq:entropySTUproduct}
S_{+} S_{-} = \frac{\pi^2\alpha^2}{4}\prod_i\left(p^i\right)^{4/3}\ \ ,
\end{equation}

\noindent
In the extremal limit we obtain the supersymmetric as well as the non-supersymmetric extremal solutions, depending on the sign chosen for the charges.

%%%%%%%%%%%%%%%%%%%%%%%%%%%%%%%%%%%%%%%%%%%%%%%%%%%%%%%%%%%%%%%%%%%%%%
%%%%%%%%%%%%%%%%%%%%%%%%%%%%%%%%%%%%%%%%%%%%%%%%%%%%%%%%%%%%%%%%%%%%%%
%%%%%%%%%%%%%%%%%%%%%%%%%%%%%%%%%%%%%%%%%%%%%%%%%%%%%%%%%%%%%%%%%%%%%%
%%%%%%%%%%%%%%%%%%%%%%%%%%%%%%%%%%%%%%%%%%%%%%%%%%%%%%%%%%%%%%%%%%%%%%

\section*{Acknowledgments}

PB and CSS wish to thank Tom\'as Ort\'in for useful discussions. CSS is happy to thank T. Mohaupt for stimulating discussions and S. Ruiz for her permanent support. CSS would like to thank the Stanford Institute for Theoretical Physics its hospitality. This work has been supported in part by the Spanish Ministry of Science and Education grant FPA2009-07692, the Comunidad de Madrid grant HEPHACOS S2009ESP-1473, and the Spanish Consolider-Ingenio 2010 program CPAN CSD2007-00042. The work of PB and CSS has been supported by the JAE-predoc grants JAEPre 2011 00452 and JAEPre 2010 00613. RD is supported by the Engineering and Physical Sciences Research Council [grant number EP/H02672X/1].

\end{document}